\documentclass[manuscript,trackchanges]{aastex701}

\begin{document}

\title{Proton Temperature Anisotropy Across Interplanetary Shocks: A Statistical Analysis with WIND observations}

\author[orcid=0000-0001-8747-7407]{Zeping Jin}
\affiliation{Department of Space Physics, University of Alabama in Huntsville, Huntsville, AL, USA}
\email{zj0016@uah.edu}  

\author[orcid=0000-0002-4299-0490]{Lingling Zhao} 
\affiliation{Department of Space Physics, University of Alabama in Huntsville, Huntsville, AL, USA}
\affiliation{Center for Space Plasma and Aeronomic Research (CSPAR), University of Alabama in Huntsville, Huntsville, AL, USA}
\email{lz0009@uah.edu}
\footnotetext[1]{Email: lz0009@uah.edu}         

\author[orcid=0000-0002-1541-6397]{Xingyu Zhu}
\affiliation{Center for Space Plasma and Aeronomic Research (CSPAR), University of Alabama in Huntsville, Huntsville, AL, USA}
\email{xz0017@uah.edu}

\author[orcid=0000-0001-5485-2872]{Vladimir Flosinski}
\affiliation{Department of Space Physics, University of Alabama in Huntsville, Huntsville, AL, USA}
\affiliation{Center for Space Plasma and Aeronomic Research (CSPAR), University of Alabama in Huntsville, Huntsville, AL, USA}
\email{zj0016@uah.edu} 

\author[orcid=0000-0002-4642-6192]{Gary P. Zank}
\affiliation{Department of Space Physics, University of Alabama in Huntsville, Huntsville, AL, USA}
\affiliation{Center for Space Plasma and Aeronomic Research (CSPAR), University of Alabama in Huntsville, Huntsville, AL, USA}
\email{zj0016@uah.edu} 

\author[orcid=0000-0001-9199-2890]{Jakobus Le Roux}
\affiliation{Department of Space Physics, University of Alabama in Huntsville, Huntsville, AL, USA}
\affiliation{Center for Space Plasma and Aeronomic Research (CSPAR), University of Alabama in Huntsville, Huntsville, AL, USA}
\email{zj0016@uah.edu} 

\author[orcid=0009-0004-4832-0895]{Yiming Jiao}
\affiliation{Department of Space Physics, University of Alabama in Huntsville, Huntsville, AL, USA}
\email{zj0016@uah.edu} 

\author[orcid=0000-0001-6286-2106]{Ashok Silwal}
\affiliation{Center for Space Plasma and Aeronomic Research (CSPAR), University of Alabama in Huntsville, Huntsville, AL, USA}
\email{zj0016@uah.edu} 

\author[orcid=0009-0008-2509-9330]{Nibuna S. M. Subashchandar}
\affiliation{Department of Space Physics, University of Alabama in Huntsville, Huntsville, AL, USA}
\email{zj0016@uah.edu} 
 
\begin{abstract}

Interplanetary (IP) shocks efficiently modify the proton temperature anisotropy of the solar wind. Analyzing $\sim$800 IP shocks observed by the Wind spacecraft from 1997–2024, we present a statistical study of upstream and downstream proton temperature anisotropy and its dependence on shock geometry, compression, and distance from the shock. We find that (1) quasi-perpendicular shocks produce a pronounced enhancement of perpendicular temperature downstream ($T_\perp > T_\parallel$), whereas parallel shocks remain near isotropic downstream due to typically stronger upstream $T_\parallel$; (2) comparisons with the Chew–Goldberger–Low (CGL) double-adiabatic model reveal geometry-dependent deviations. CGL overestimates downstream perpendicular heating and underestimates parallel heating at quasi-perpendicular shocks, with the opposite trend at quasi-parallel shocks, highlighting the importance of non-adiabatic processes beyond simple compression; (3) Shock-driven anisotropy is strongly localized near the shock and gradually relaxes toward typical solar wind conditions farther downstream as the shock’s influence diminishes; and (4) downstream anisotropy is regulated by kinetic instabilities, with quasi-perpendicular shocks constrained by proton cyclotron and mirror instabilities and quasi-parallel shocks limited by the parallel firehose instability. Together, these results show that the evolution of temperature anisotropy at interplanetary shocks is controlled by shock geometry, localized processes, and instability driven regulation.

\end{abstract}



\section{Introduction}

Collisionless shocks are ubiquitous in space plasmas and play a central role in plasma heating, particle acceleration, and energy dissipation \citep{2009A&ARv..17..409T,2017RvMPP...1....1P}. In the heliosphere, interplanetary (IP) shocks driven by coronal mass ejections or stream interaction regions provide a natural laboratory for studying shock physics \citep{zhao2019particle, zhao2021turbulence}. In situ spacecraft measurements enable direct access to plasma moments and magnetic fields, allowing detailed tests of both fluid and kinetic theories \citep{2011JGRA..11610105H,2022FrASS...905672T, Zhao_2018, zhao2021mhd}.
A key kinetic signature of collisionless shocks is the development of temperature anisotropy relative to the background magnetic field. Both upstream and downstream proton populations frequently show $T_\perp \ne T_\parallel$, reflecting the combined influence of compression, magnetic field amplification, wave–particle interactions, and kinetic instabilities {\citep{2009A&ARv..17..409T, Burgess2015, 2002GeoRL..29.1839K, 2006GeoRL..33.9101H}}. Theoretical treatments incorporating anisotropic pressure, including anisotropic Rankine–Hugoniot relations, show that shock jump conditions depend sensitively on parallel and perpendicular temperatures \citep{1970P&SS...18.1611H,1986JGR....91.6771L,1995GeoRL..22.2409C,2000JPlPh..64..561E,2009ASTRA...5...31G}. The Chew–Goldberger–Low (CGL) double-adiabatic theory \citep{1956RSPSA.236..112C} provides a baseline description of anisotropic plasma evolution, {but it generally cannot reproduce observed solar wind features \citep{2019LRSP...16....5V}. Its applicability at shocks is further limited by non-adiabatic processes, including cross-shock potentials, shock drift, and wave–particle interactions \citep{1984JGR....89.6654G,1988SSRv...48..195D,2009A&ARv..17..409T}. }

Temperature anisotropy is also closely tied to kinetic instabilities that regulate plasma states. In the solar wind, proton anisotropies are frequently observed near the marginal thresholds of mirror and firehose instabilities \citep{2002GeoRL..29.1839K,2006GeoRL..33.9101H}, indicating that these microinstabilities limit departures from isotropy. Theory and simulations further show how streaming and damping effects can modify these thresholds \citep{2011PhRvL.107t1102S,2018A&A...613A..23V}. Shocks themselves generate anisotropies that drive downstream instabilities, whose nature depends strongly on shock geometry: quasi-perpendicular shocks preferentially excite mirror and ion cyclotron modes \citep{1970JGR....75.4666K,1998JGR...10311961B,2007ApJ...659L..65L}, whereas quasi-parallel shocks favor the parallel firehose instability \citep{1994JGR....99.5877A,2001AnGeo..19..275C,2007ApJ...659L..65L,2011PhPl...18f2110L,2022MNRAS.509.2084H}.

In this work, we present a statistical study of $\sim$800 IP shocks observed by Wind from 1997–2024, examining how upstream and downstream proton temperature anisotropy depends on shock parameters such as obliquity and compression ratio. We further compare the observations with CGL double-adiabatic predictions and evaluate the role of kinetic instabilities in regulating downstream anisotropy. This comprehensive approach allows us to quantify the geometry-dependent modification of proton temperature by shocks, assess the limitations of fluid models, and clarify how instabilities shape the downstream plasma. The paper is organized as follows. Section~\ref{sec:data} describes the data and event selection. Section~\ref{sec:Ad} presents the dependence of temperature anisotropy on shock parameters and compares the observations with CGL predictions. Section~\ref{sec:kine} examines kinetic instabilities upstream and downstream, and Section~\ref{sec:Conclution} summarizes the main conclusions and discussion.

\section{Data Selection}\label{sec:data}
We use in situ observations from the \textit{Wind} spacecraft covering the period 1997–2024. Interplanetary shock events are taken from the Database of Interplanetary Shocks maintained at ipshocks.fi\footnote{\url{https://ipshocks.fi}}, in which shock normals, density compression ratios, and shock obliquities are determined using 8-minute averaged plasma and magnetic field measurements upstream and downstream of the shock \citep{kilpua2015properties}. The data set includes approximately 800 interplanetary shocks, encompassing both fast-forward (FF) and fast-reverse (FR) shocks. 
{We note that automatic shock detection can vary between catalogs. However, because our study analyzes a large sample of shocks, the reported statistical trends are unlikely to be influenced by the inclusion or exclusion of a few events. We also acknowledge that other shock catalogs exist (e.g., the Wind shock database at CfA\footnote{\url{https://lweb.cfa.harvard.edu/shocks/wi_data/index.html}}), and cross-comparisons may be valuable for future studies.}
Proton temperature anisotropy is obtained from the Wind/3DP instrument \citep{lin1995three}. We use the 24-second resolution Ion OmniDirectional Fluxes and Moments data product {(WI\_PLSP\_3DP)}, which provides the proton temperature components derived from measured ion velocity distribution functions \citep{WIPLSP3DP}. {This dataset provides separate temperature tensors for protons and alpha particles, minimizing contamination from alpha particles. We use the proton temperature anisotropy for our statistical analysis.}
The proton temperature tensor is rotated into a magnetic field-aligned coordinate system using the local magnetic field direction measured by the Wind/MFI instrument \citep{lepping1995wind}. The parallel temperature $T_{\parallel}$ is taken as the tensor component along the magnetic field direction, while the perpendicular temperature $T_{\perp}$ is obtained by averaging the two diagonal components perpendicular to the magnetic field. 
Off-diagonal elements of the temperature tensor are not included in the calculation of temperature anisotropy. 

{To ensure consistency with the used shock database, the temperature anisotropy, defined as $A = T_\perp / T_\parallel$, is calculated by averaging the perpendicular ($T_\perp$) and parallel ($T_\parallel$) proton temperatures within an 8-minute window upstream and downstream of each shock when analyzing its dependence on shock properties such as compression ratio and obliquity. To assess how anisotropy evolves with distance from the shock, $A$ is also computed using multiple averaging windows at different times relative to the shock crossing, allowing us to distinguish shock-localized anisotropy enhancements from the more relaxed solar wind farther away. To minimize potential saturation effects in 3DP measurements near interplanetary shocks \citep{2010PhDT........43W}, plasma data within one minute of the shock ramp are excluded from the analysis.}

In the following section, we present a statistical analysis of proton temperature anisotropy across interplanetary shocks, including its dependence on shock geometry and compression ratio, its comparison with predictions from the CGL double-adiabatic theory, and the role of kinetic instabilities in regulating the downstream plasma state.

\section{Results: Temperature Anisotropy Across Shocks} \label{sec:Ad}
\subsection{Dependence on Shock Obliquity} \label{sec:obliquity}

We examine the proton temperature anisotropy $A$, in the vicinity of approximately 800 interplanetary shocks observed by the \textit{Wind} spacecraft. As described above, the 24-s resolution temperature components data are averaged over 8-minute intervals upstream and downstream of each shock in order to remain consistent with the shock parameter determinations. 


\begin{figure}[htbp]
    \centering
    \includegraphics[width=\linewidth]{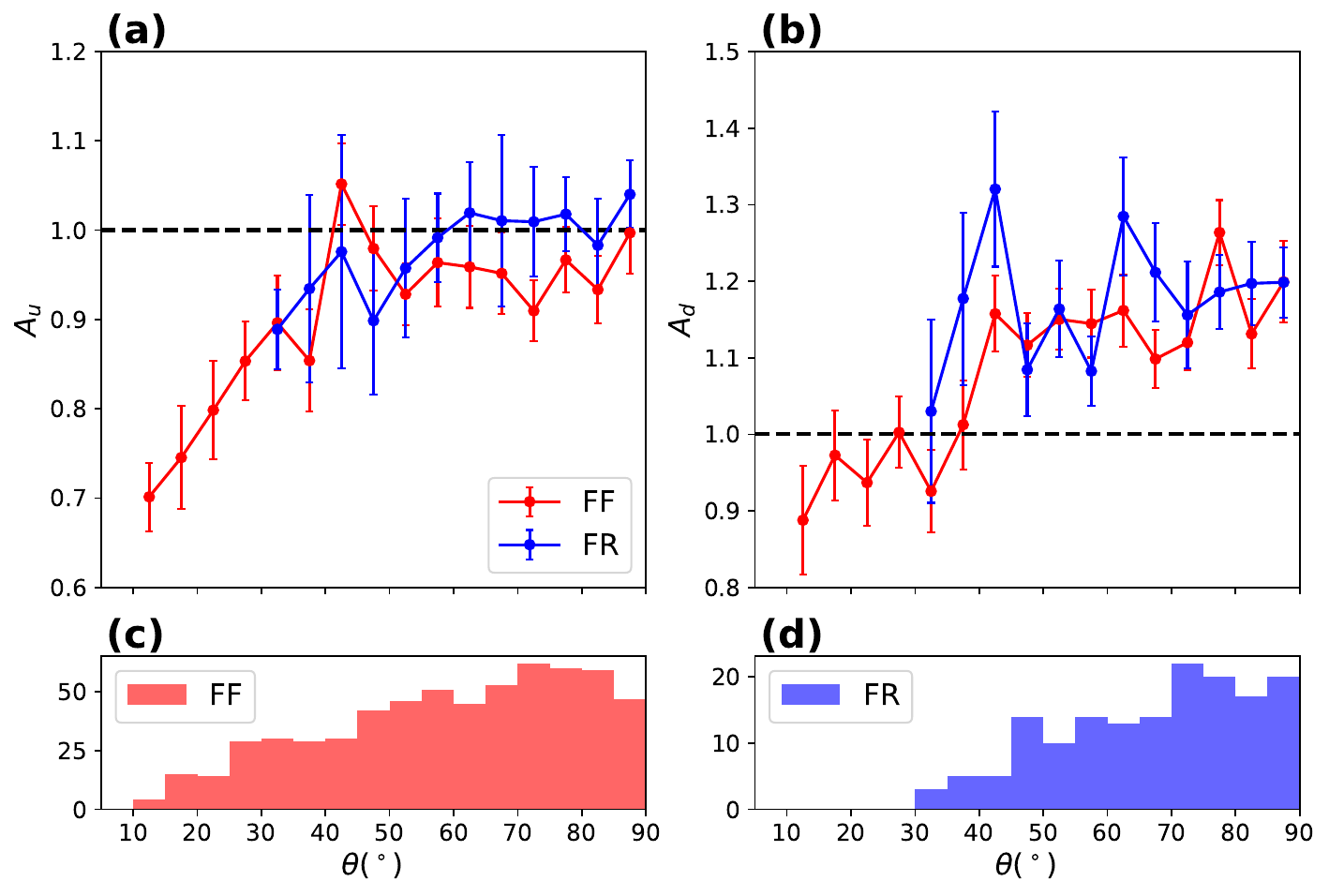} 
    \caption{
Upstream $A_\mathrm{u}$ (left panel) and downstream $A_\mathrm{d}$ (right panel) proton temperature anisotropy, defined as $A = T_\perp/T_\parallel$, as a function of shock obliquity $\theta\,(^\circ)$. Fast-forward (FF) shocks are shown in red and fast-reverse (FR) shocks in blue. Each dot represents the binned mean anisotropy within a $5^\circ$ shock-obliquity bin, and {the error bar}
denote the standard error of the mean. {The lower two panels show the number distribution of FF shocks (red) and FR shocks (blue) in each $\theta$ bin corresponding to the upper panels.}
}
\label{fig:Para1}
\end{figure}

Figure~\ref{fig:Para1} illustrates the dependence of temperature anisotropy on shock obliquity $\theta \,(^\circ)$. Panels (a) and (b) show the upstream ($A_\mathrm{u}$) and downstream ($A_\mathrm{d}$) anisotropy, respectively. Fast-forward (FF) shocks are shown in red, while fast-reverse (FR) shocks are shown in blue. Most FR shocks observed are quasi-perpendicular.
To highlight systematic trends, the solid lines with dots represent binned averages of $A$ calculated over 16 shock-obliquity bins, each with a width of $5^\circ$. The {error bar}
denote the standard error of the mean within each bin, providing a measure of the statistical uncertainty.
{We have also examined the dependence of temperature anisotropy on other shock parameters, including Alfv\'en Mach number, shock speed, and upstream plasma $\beta$ (not shown), but no significant systematic trends are observed among these parameters.} Shock obliquity emerges as the dominant factor controlling the variation of temperature anisotropy across shocks. For quasi-parallel shocks, the upstream plasma exhibits $A_u < 1$, as shown in Figure~\ref{fig:Para1}(a), indicating that the parallel temperature $T_\parallel$ significantly exceeds the perpendicular temperature $T_\perp$. In contrast, for quasi-perpendicular shocks, the upstream plasma is nearly isotropic, with $A_u \simeq 1$, particularly for fast-reverse shocks. The stronger dominance of $T_\parallel$ upstream of quasi-parallel shocks likely arises from enhanced field-aligned ion populations and upstream wave activity, which preferentially broaden the parallel velocity distribution, while perpendicular heating remains weak in the absence of strong magnetic-field compression {\citep{burgess2005quasi, eastwood2005foreshock, 2013JGRA..118..823O}}.

Downstream of the shock shown in Figure~\ref{fig:Para1}(b), the temperature anisotropy $A_d$ increases for both shock geometries. For quasi-parallel shocks, the downstream plasma becomes nearly isotropic, with $A_d \sim 1$, indicating that perpendicular heating is substantially enhanced but still insufficient to fully isotropize the plasma, as $A_d$ remains slightly below unity. In contrast, quasi-perpendicular shocks exhibit a much stronger downstream anisotropy, with $A_d > 1$ and $T_\perp > T_\parallel$. This behavior is expected due to the significant magnetic-field compression across quasi-perpendicular shocks, which directly enhances perpendicular heating through adiabatic and kinetic processes \citep[e.g.,][]{1986JGR....91.6771L}. In general, FR shocks exhibit similar upstream and downstream anisotropy characteristics as quasi-perpendicular FF shocks, with nearly isotropic upstream plasma and enhanced perpendicular heating downstream.


\subsection{Comparison with CGL Predictions} \label{sec:CGL}

The Chew-Goldberger-Low (CGL) double-adiabatic theory \citep{chew1956boltzmann} provides a useful framework for understanding the evolution of plasma temperature anisotropy under the assumption of conservation of the first and second adiabatic invariants. 
{The perpendicular temperature evolves with the magnetic field as $T_\perp \propto B$ due to the conservation of the magnetic moment, while the parallel temperature evolves as $T_{||} \propto n^2/B^2$ due to the conservation of the longitudinal invariant.}
{While the CGL model describes the general trend expected from adiabatic invariants, it is well known that it fails to reproduce many features of the solar wind observation\citep[e.g.,][]{verscharen2016collisionless}. Observed plasma frequently deviates from CGL predictions due to kinetic effects, including wave–particle interactions and non-adiabatic processes \citep[e.g.,][]{2018PhLA..382.2052K, 2022MNRAS.509.2084H}. Motivated by this, we systematically compare observations near interplanetary shocks with CGL predictions to assess where additional physical processes, such as shock-driven heating, wave–particle interactions, or kinetic instabilities, become important.} 

\begin{figure}[htbp]
    \centering
    \includegraphics[width=\linewidth]{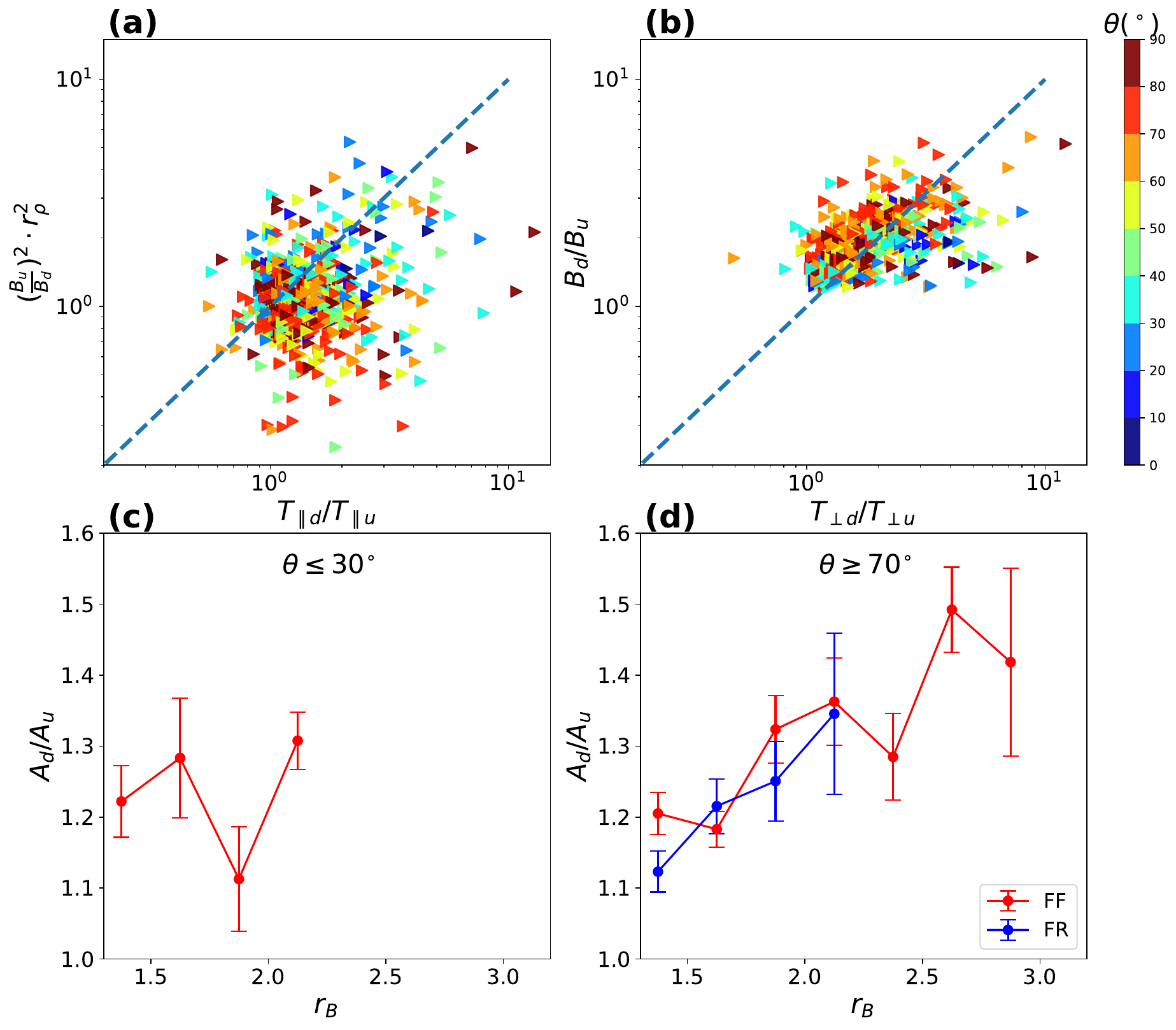} 
    \caption{Comparison of observed downstream-to-upstream proton temperature ratios with CGL predictions. Panels (a) and (b) show the parallel ($T_{\parallel d}/T_{\parallel u}$) and perpendicular ($T_{\perp d}/T_{\perp u}$) temperature ratios, respectively, with markers color-coded by shock obliquity. The dashed line denotes the temperature ratio predicted by the CGL double-adiabatic model based on magnetic-field and density compression. Deviations from this line indicate the influence of non-adiabatic and kinetic processes across the shock. Panels (c) and (d) show the relative change in temperature anisotropy, $A_d/A_u$, as a function of the magnetic compression ratio $r_B$ for quasi-parallel (c) and quasi-perpendicular (d) shocks, respectively. 
    }
    \label{fig:CGL}
\end{figure}

Figure~\ref{fig:CGL}(a) and (b) compare the observed downstream-to-upstream parallel and perpendicular temperature ratios, $T_{\parallel d}/T_{\parallel u}$ and $T_{\perp d}/T_{\perp u}$, with the predictions of the CGL double-adiabatic model. Panels (a) and (b) only consider fast-forward shocks, which span a wider range of shock obliquities. For the parallel temperature, CGL predicts that the temperature change results from the combined effects of magnetic field and density compression, scaling as $T_{\parallel d}/T_{\parallel u} \propto (B_u/B_d)^2(n_d/n_u)^2$. For the perpendicular component, the model predicts linear scaling with magnetic compression, $T_{\perp d}/T_{\perp u} \propto B_d/B_u$. These theoretical relations are indicated by the blue dashed lines in panels (a) and (b). Data points above the reference lines indicate that CGL overestimates the observed temperature increase, while data below the line indicate that CGL underestimates the temperature change.

Our comparison reveals systematic deviations from CGL that depend on shock geometry. For quasi-perpendicular shocks, the downstream perpendicular temperature increases less than predicted by adiabatic compression (panel (b)), whereas the parallel temperature is underestimated by CGL (panel (a)). This suggests that magnetic moment conservation is partially violated at these shocks, likely due to strong magnetic gradients, cross-shock electric fields, and kinetic processes that redistribute energy from perpendicular to parallel directions. In contrast, for quasi-parallel shocks, the perpendicular temperature is slightly larger than the CGL prediction, while the parallel temperature is broadly consistent with CGL expectations. This implies that wave–particle interactions and pitch-angle scattering provide additional perpendicular heating beyond adiabatic compression for quasi-parallel shocks. 

In Figure~\ref{fig:CGL}(c) and (d), we present the relative change in temperature anisotropy across the shock, $A_d/A_u$, for quasi-parallel ($\theta \le30^\circ$) and quasi-perpendicular ($\theta \ge 70^\circ$) shocks, respectively, as a function of the shock compression ratio computed from magnetic compression. {We chose $30^\circ$ and $70^\circ$ as thresholds mainly to account for the uncertainty in the shock normal estimation from single-spacecraft data, avoiding cases with oblique angles near the parallel or perpendicular boundaries. In addition, Figure~\ref{fig:Para1} shows that the transition in downstream anisotropy $A_d$ occurs near $\theta \sim 40 ^\circ$, with more parallel shocks having $A_d <1$ and more oblique shocks having $A_d >1$. We test using a $45^\circ$ threshold to distinguish quasi-parallel and quasi-perpendicular shocks, and the resulting trends do not differ significantly from those shown in Figure~\ref{fig:CGL}(c) and (d).} On average, $A_d > A_u$ is observed for nearly all shocks, indicating that perpendicular heating is relatively more efficiently enhanced across the shock. In the CGL limit, for quasi-parallel shocks where magnetic compression is weak, the temperature anisotropy is expected to decrease with increasing compression ratio, following $A_d/A_u \propto r_{\rm sh}^{-2}$. In contrast, for quasi-perpendicular shocks, where the magnetic compression is comparable to the density compression, the CGL model predicts $A_d/A_u \propto r_{\rm sh}$. However, panel (c) shows that $A_d/A_u$ for quasi-parallel shocks decreases much more slowly than the CGL prediction and even increases at strong shocks. This behavior is consistent with additional non-adiabatic perpendicular heating not captured by the CGL model, leading to a weaker reduction or even enhancement of temperature anisotropy, as also suggested by the deviations seen in panels (a) and (b).
For quasi-perpendicular shocks, $A_d/A_u$ increases with compression ratio but at a significantly slower rate than the CGL prediction of $A_d/A_u \propto r_{\rm sh}$. This behavior is also consistent with the CGL model overestimating perpendicular heating and underestimating parallel heating for perpendicular shocks, as discussed in panels (a) and (b). We further find no significant difference between fast–forward and fast–reverse shocks in terms of the dependence on compression ratio. We note that the number of quasi-parallel shocks in our dataset is relatively small. Therefore, the trends observed for quasi-parallel shocks should be interpreted with caution.

{We emphasize that the CGL model is not expected to hold at collisionless shocks, where particle reflection, wave–particle interactions, and dissipation violate its assumptions. Nevertheless, it provides a baseline for the temperature anisotropy changes that would result from ideal adiabatic compression of the plasma. Deviations from the CGL predictions, therefore, serve as a quantitative diagnostic of non-adiabatic heating and its directional dependence.} In this sense, the observed statistical excesses or deficits of $T_{\parallel d}/T_{\parallel u}$ and $T_{\perp d}/T_{\perp u}$ relative to CGL predictions reflect the influence of kinetic effects or large-scale shock dynamics, which vary systematically with shock obliquity.

\subsection{Effect of Distance from the Shock} \label{sec:distance}

\begin{figure}[htbp]
    \centering
    \includegraphics[width=\linewidth]{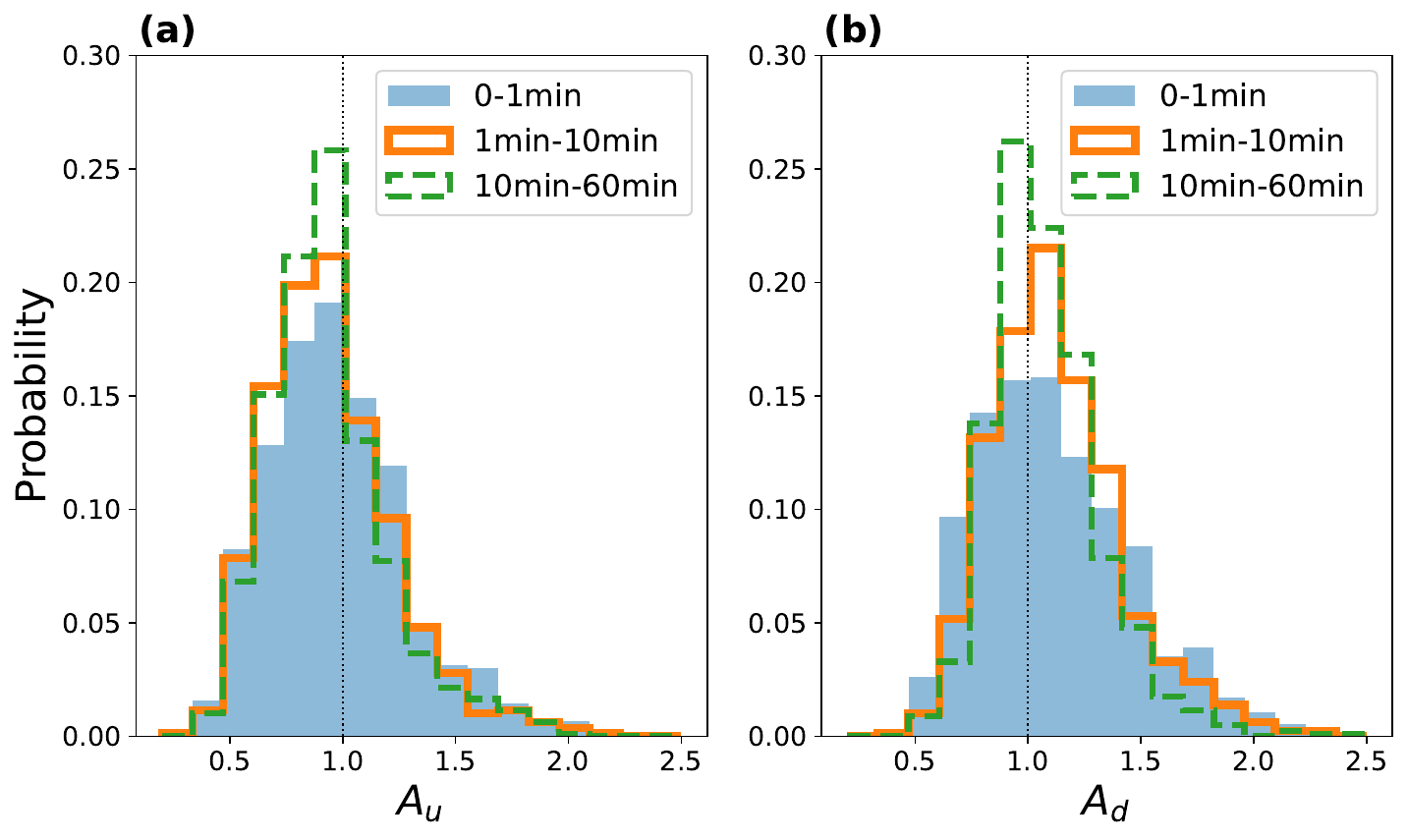} 
    \caption{Probability distributions of proton temperature anisotropy $A$ at different time intervals relative to the shock front. The left and right panels show upstream and downstream distributions, respectively. Distributions are computed for increasing time windows from the shock: blue (0–1~min), orange (1–10~min), and green dashed (10–60~min). Shorter intervals reveal stronger anisotropy near the shock, while longer intervals approach more isotropic solar wind conditions.}
    \label{fig:dist}
\end{figure}

To investigate how shocks modify the solar wind temperature anisotropy, we examine
$A = T_\perp/T_\parallel$ at varying distances relative to the shock crossing for all FF shocks. 
Figure~\ref{fig:dist} shows probability distributions of upstream {($A_u$, in the left panel)} and downstream {($A_d$, in the right panel)} proton temperature anisotropy 
for three representative time intervals relative to the shock front: 0-1~min, 1–10~min, and 10–60~min. 
{$A$ is computed by averaging the measured $T_\perp$ and $T_\parallel$ within these specified time windows relative to the shock. In Figure 3, the 0–1~min and 1–10~min intervals represent plasma conditions close to the shock, while the 10–60~min interval serves as a baseline representing ambient solar wind conditions sufficiently far from the shock. {Measurements within one minute of the shock transition are excluded from the analysis to minimize potential detector saturation effects near the shock ramp (see Section~\ref{sec:data}).} These intervals allow us to examine how the statistical properties of temperature anisotropy evolve with increasing temporal distance from the shock. We note that the overall trends in temperature anisotropy evolution are consistent when using alternative time interval lengths.
}

For both upstream and downstream plasmas, the distributions closest to the shock (0-1~min) exhibit a reduced probability of isotropy ($A \approx 1$), particularly downstream, indicating enhanced anisotropy in the immediate shock vicinity. 
{As the averaging interval moves farther from the shock (orange and green curves), the probability of isotropic plasmas increases, indicating a gradual relaxation toward typical solar wind conditions. For the farthest interval considered here (10–60~min), both upstream and downstream temperature anisotropy distributions exhibit a much higher probability near the peak at $A \sim 1$ compared with the other two closer intervals. In contrast, the downstream 0–1 min and 1–10 min intervals have peaks that are clearly shifted to the right and are more skewed toward $A_d > 1$, indicating stronger perpendicular
temperature anisotropy (i.e., larger $A_d$ values) closer to the shock in the downstream region. The upstream difference
shown in Figure~\ref{fig:dist}(a) is less pronounced than in the downstream case. However, the 10–60~min interval remains more narrowly distributed and concentrated near $A_u \sim 1$, while the shorter intervals are slightly broader and show a bit higher probability of anisotropic states. Overall, both the
upstream and downstream trends suggest that the shock-related temperature anisotropy gradually
weakens with increasing distance from the shock.} 

{Although the plasma observed farther from the shock may not correspond to the same parcel that previously crossed it, Figure \ref{fig:dist} is intended to characterize how temperature anisotropy changes with temporal distance from the shock upstream and downstream, respectively. The 10–60~min intervals are compared only with their respective upstream or downstream intervals closer to the shock front (0–1~min and 1–10~min) to illustrate the relaxation of anisotropy with distance from the shock.}

The distributions in Figure \ref{fig:dist} also confirm that most upstream plasmas are characterized by $A_u < 1$, while downstream plasmas have $A_d > 1$, consistent with the perpendicular heating enhancement across shocks discussed in Figure~\ref{fig:Para1}. These results provide clear evidence that interplanetary shocks effectively drive changes in proton temperature anisotropy, with the strongest effects occurring in the immediate shock vicinity.


\section{Results: Kinetic Instabilities Across Shocks}\label{sec:kine}

\begin{figure}[htbp]
    \centering
    \includegraphics[width=\linewidth]{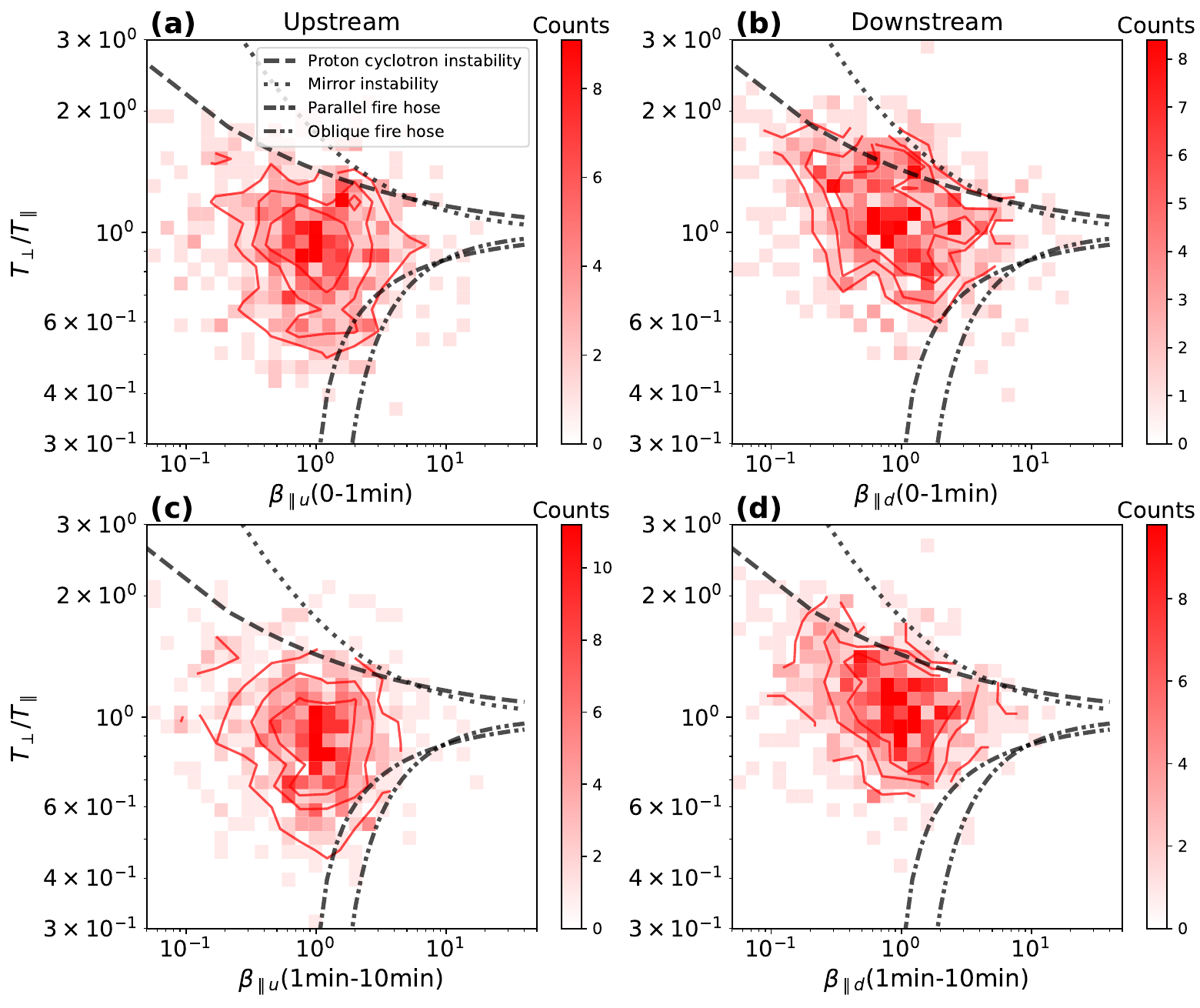} 
    \caption{Statistical distribution of ($T_\perp/T_\parallel$, $\beta_\parallel$) upstream and downstream of Fast-Forward shocks. {The black Dashed or Dotted} lines indicate theoretical thresholds for various instabilities with growth rate $\gamma = 10^{-3}\,\omega_{cp}$. {(Dotted: Mirror instability; Dashed: Proton cyclotron instability; Dash-dot: Parallel fire hose instability; Dash-dot-dot: Oblique fire hose instability)}
    The top panels show the 0–1~min interval immediately near the shock, while the bottom panels show the 1–10~min interval farther from the shock. These time windows correspond to those used in Figure~\ref{fig:dist}. {{Red} contours represent the density distribution, and its levels span from 10\% to 50\% of the maximum value with equal spacing.}
    }
    \label{fig:Ins1_FF}
\end{figure}


Our analysis above shows that interplanetary shocks enhance the perpendicular temperature downstream and produce systematic deviations from CGL predictions, indicating that processes beyond simple adiabatic compression influence the plasma state. In collisionless shocks, such strong temperature anisotropies can drive kinetic instabilities, which act to limit the further growth of anisotropy. Motivated by this, we examine plasma instabilities in both the upstream and downstream regions. Instead of focusing on individual events to identify specific wave modes \citep{zhao2026}, we adopt a statistical approach, which enables us to capture the systematic and common characteristics of shock-related instabilities across a large sample of events.

\subsection{Temporal Evolution of Instabilities} \label{sec:inst_time}
\begin{figure}[htbp]
    \centering
    \includegraphics[width=\linewidth]{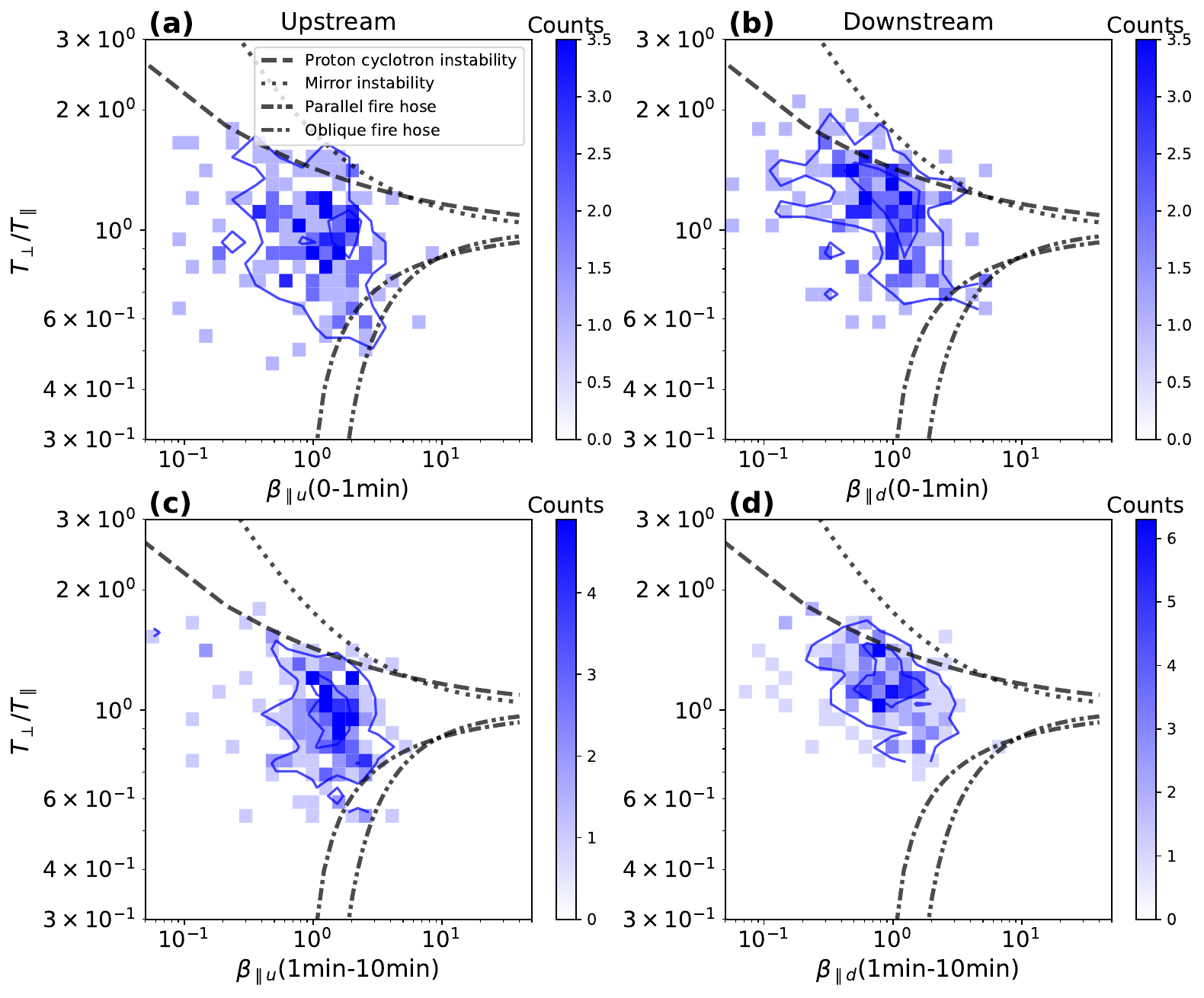} 
    \caption{Similar to Figure~\ref{fig:Ins1_FF}, but showing the instability distributions for Fast-Reverse (FR) shocks.}
    \label{fig:Ins1_FR}
\end{figure}

Figures~\ref{fig:Ins1_FF} 
and \ref{fig:Ins1_FR} 
show the statistical distributions of ($T_\perp/T_\parallel$, $\beta_\parallel$) for fast-forward and fast-reverse shocks, respectively. To probe the effect of distance from the shock, we use two representative time windows: {a closer interval (0-1~min) capturing conditions immediately adjacent to the shock, and a farther interval (1–10~min) sampling regions more distant from the shock.}
{These two cases correspond to the blue and orange groups in Figure~\ref{fig:dist}.}
For both shock types, {the upstream plasma is generally less anisotropic than the downstream plasma, and the instability thresholds are therefore less populated upstream. This is evident in Figure~\ref{fig:Ins1_FF}(a) and (c), where most events upstream plasma cluster near the nearly isotropic state and few events approach the instability boundaries. In contrast, comparing Figures~\ref{fig:Ins1_FF}(a) and (b) or (c) and (d) reveals a clear transition: the high-count regions (deep red) shift from the center toward the instability boundaries downstream, especially evident at low plasma $\beta$. As a result, the downstream plasma exhibits stronger constraints from the proton cyclotron, mirror, and parallel firehose instabilities. The upstream distributions show less well-defined instability boundaries.} 
{Immediately downstream of the shock (panel b), a larger fraction of events approach the mirror-instability boundary, consistent with strong perpendicular heating near the shock ramp \citep[e.g.,][]{czaykowska1998mirror, burgess2007shock}. Farther downstream (panel d), the plasma anisotropy becomes increasingly constrained by the proton cyclotron instability. This trend suggests that wave–particle interactions gradually regulate the perpendicular anisotropy as the plasma evolves away from the shock. We note that this describes a statistical trend, and individual events may deviate from this behavior. A similar pattern is observed for FR shocks (Figure~\ref{fig:Ins1_FR}), although the available database for FR shocks is limited.}

These statistical distributions also help explain the nonlinear behavior observed between downstream and upstream anisotropy (not shown). For small to moderate upstream anisotropy, the downstream anisotropy increases systematically. However, for large upstream anisotropy, the proton cyclotron and mirror instabilities impose upper bounds on $A_d$, limiting its further growth. As a result, $A_d$ saturates or increases more slowly with $A_u$ at high upstream anisotropy. This indicates that kinetic instabilities act as a regulatory mechanism, constraining extreme downstream anisotropy and explaining why the relationship between $A_d$ and $A_u$ deviates from strict linearity for strongly anisotropic upstream plasmas.

\subsection{Shock Obliquity Dependence of Instabilities} \label{sec:inst_angle}
\begin{figure}[htbp]
    \centering
    \includegraphics[width=\linewidth]{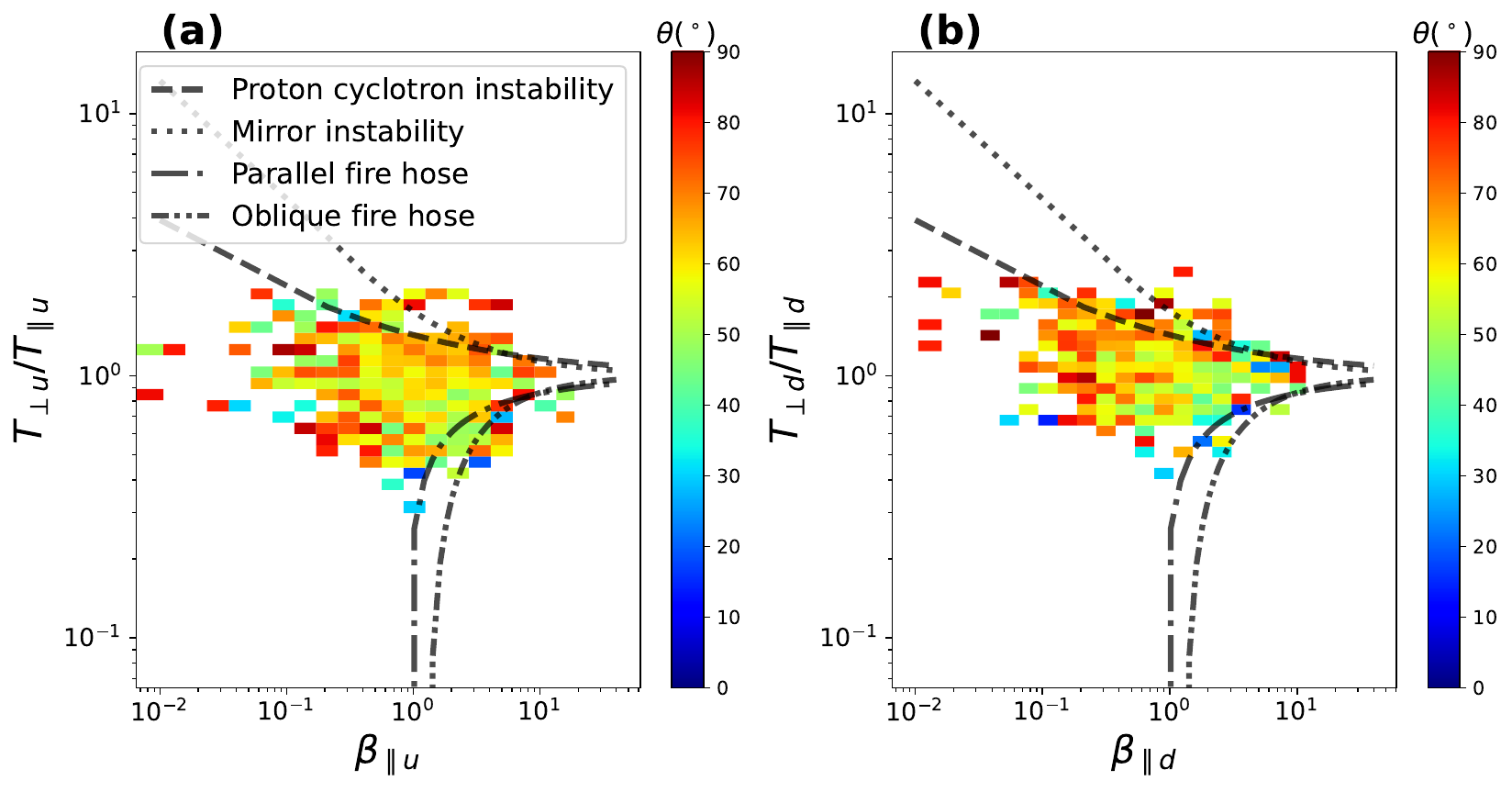} 
    \caption{Shock-angle-resolved distributions of ($T_\perp/T_\parallel$, $\beta_\parallel$) for FF shocks, using an 8-minute averaging window upstream and downstream, which matches the window used to calculate the shock angle. The left and right panels show upstream and downstream distributions, respectively, and the color indicates the average shock obliquity in each bin.}
    \label{fig:Ins2}
\end{figure}

Finally, Figure~\ref{fig:Ins2} illustrates how the dominant instabilities vary with shock obliquity. {Here, we use an 8-minute window upstream and downstream, matching the window used to calculate the shock angle. Proton temperature anisotropy and beta are calculated as the average values within 8-minute interval. For the upstream distributions shown in Figure~\ref{fig:Ins2}(a), no clear trend is observed between instability development and shock angle. In the downstream region (panel b), the distribution of points (blue/cyan) relative to the parallel firehose threshold is statistically significant for quasi-parallel shocks, although some individual events scatter near other instability regions, such as the proton cyclotron instability.}

{This trend is also reflected in Figure~\ref{fig:Para1}. For quasi-parallel shocks, the downstream temperature on average satisfies $T_\perp < T_\parallel$, consistent with the firehose instability. However, some individual parallel shocks exhibit $T_\perp/T_\parallel \ge 1$ downstream, particularly at slightly higher or near-oblique shock angles, though these events are not statistically significant. Due to uncertainties in the shock normal estimation, these ``near-oblique'' cases may also not represent strictly parallel shocks. In contrast, quasi-perpendicular downstream plasma typically exhibits $T_\perp > T_\parallel$ and clusters near the proton cyclotron instability boundary for the intermediate interval considered in Figure~\ref{fig:Ins2}, reflecting enhanced perpendicular heating downstream of perpendicular shocks.}

{We also tested different windows for Figure~\ref{fig:Ins2}, and the overall trends remain unchanged. Comparing Figure~\ref{fig:Ins2}(b) with Figure~4(b), using a shorter interval (closer to the shock) results in slightly more perpendicular shock events downstream being constrained by the mirror instability. We emphasize that these results represent statistical trends across a large sample of shocks. The detailed behavior of individual events warrants further investigation \citep[e.g.,][]{zhao2026}.}

{The effect of shocks on temperature anisotropy is strongly geometry-dependent. Perpendicular shocks drive stronger perpendicular heating, bringing the plasma closer to the proton cyclotron instability at intermediate distances from the shock, or to the mirror instability in the immediate downstream region. In contrast, parallel shocks exhibit weaker perpendicular heating and are generally more constrained by the firehose instability. These results confirm that the regulation of anisotropy by kinetic instabilities depends on both the distance from the shock and the shock geometry.}


\section{Summary and Discussion} \label{sec:Conclution}

We present a statistical analysis of proton temperature anisotropy across interplanetary shocks observed by Wind, examining the roles of shock obliquity, compression, distance from the shock, and kinetic instability constraints. By comparing observations with CGL double-adiabatic predictions and instability thresholds, we quantify how collisionless shocks generate and regulate proton anisotropy in the solar wind. Our main findings are summarized as follows.

1. Shock obliquity is the dominant factor controlling temperature anisotropy changes near the Shock. Upstream of quasi-parallel shocks the plasma typically exhibits $A_u = T_{\perp u}/T_{\parallel u}<1$, while upstream of quasi-perpendicular shocks it is close to isotropic. Downstream of both shock types, perpendicular temperature is enhanced, but quasi-perpendicular shocks produce substantially larger $A_d>1$. This is consistent with stronger magnetic compression and perpendicular energization at perpendicular shocks \citep[e.g.,][]{burgess2007shock,schwartz2022energy, zhao2025theory, howes2025velocity}.

2. Downstream temperatures systematically deviate from CGL predictions.
While the CGL model captures some qualitative trends, it fails to reproduce the observed downstream temperature partition. For quasi-perpendicular shocks, CGL overestimates $T_{\perp d}$ and underestimates $T_{\parallel d}$, indicating partial violation of adiabatic invariance due to cross-shock electric fields and kinetic processes \citep[e.g.,][]{hull2012multiscale,gedalin2024electron}. For quasi-parallel shocks, $T_{\parallel}$ remains closer to CGL expectations, while $T_{\perp}$ is enhanced beyond adiabatic predictions, suggesting additional perpendicular heating from wave–particle scattering.

3. {Shock-driven anisotropy decreases with distance from the shock. Immediately downstream, the plasma exhibits enhanced temperature anisotropy that depends on shock geometry. This anisotropy is primarily generated by shock-mediated processes, rather than by the background solar wind or large-scale adiabatic effects. At farther distances, the plasma gradually relaxes toward the background solar wind state, although some residual shock-driven anisotropy persists. } 

4. {Kinetic instabilities regulate downstream anisotropy in a geometry-dependent manner. For parallel shocks, the downstream plasma is generally constrained by the parallel firehose instability, reflecting slightly stronger parallel heating, although some individual events may also be influenced by the proton cyclotron instability. In contrast, for quasi-perpendicular shocks, the plasma approaches the mirror instability in the immediate downstream region and the proton cyclotron instability at larger distances from the shock.} 

{A recent study by \cite{2019A&A...632A..92D} using Wind/3DP data found that the perpendicular temperature $T_\perp$ in fast solar wind is enhanced at large angles between the magnetic field and the radial direction ($\theta_{BR}$), which they suggested that this trend is likely related to proton cyclotron resonance. Instrumental effects were ruled out via simulated sampling of proton velocity distribution functions (VDFs). In our study, $T_\perp$ enhancement occurs primarily downstream of quasi-perpendicular shocks due to shock-driven heating and wave–particle interactions. We suggest that the positive correlation between $T_\perp$ and $\theta_{BR}$ in fast wind reported by \cite{2019A&A...632A..92D} may partly reflect the presence of shocks in their dataset, downstream of which fast solar wind streams are typically observed. 
However, we note that Wind/3DP 24-second temperature components data should be interpreted cautiously, as the methods used to derive and average plasma moments from the measured 3D VDFs may influence the resulting temperature anisotropy. Nevertheless, the statistical trends presented here are based on a large shock sample and therefore reflect systematic behavior. Future studies using Parker Solar Probe and Solar Orbiter, which provide higher-resolution plasma measurements closer to the Sun, will help further test these results.}

Overall, our study demonstrates that proton anisotropy downstream of interplanetary shocks is determined by the combined effects of shock geometry, non-adiabatic heating processes, and kinetic instability constraints. While the smaller number of quasi-parallel shocks introduces larger statistical uncertainty for that subset, the observed trends are generally consistent with previous studies of collisionless shock heating and anisotropy regulation. Our results provide new observational constraints for models of solar wind heating and energetic particle transport at collisionless interplanetary shocks. 

\begin{acknowledgments}
{The authors thank the reviewer for helpful comments.} We acknowledge the partial support of the NSF EPSCoR RII-Track-1 Cooperative Agreement OIA2148653, NSF award 2442628, NASA awards 80NSSC20K1783, 80NSSC21K1319, 80NSSC23K0415, 80NSSC24K1867, 80NSSC25K7750. We acknowledge the use of WIND spacecraft data, publicly
available at NASA/SPDF (\url{https://cdaweb.gsfc.nasa.gov}). We
also acknowledge the use of the Interplanetary Shock Database maintained at the University of Helsinki (\url{http://www.ipshocks.fi}).
\end{acknowledgments}




\bibliography{TA}{}
\bibliographystyle{aasjournalv7}

\end{document}